\documentclass[12pt]{article}

\usepackage{newtxtext,newtxmath}
\usepackage{graphicx}

\usepackage[letterpaper,margin=1in]{geometry}

\renewenvironment{abstract}
	{\quotation}
	{\endquotation}

\date{}

\makeatletter
\renewcommand{\fnum@figure}{\textbf{Figure \thefigure}}
\renewcommand{\fnum@table}{\textbf{Table \thetable}}
\makeatother

\usepackage{scicite}

\usepackage{url}

\def\scititle{
	Bar-driven secular evolution largely complete in a disk galaxy 7.6 billion years ago
}
\title{\bfseries \boldmath \scititle}

\author{
    Takafumi~Tsukui$^{1,2,3\ast}$,
    Zeyu~Gao$^{1,4,5}$,
    Emily~Wisnioski$^{3}$,
    Joss~Bland-Hawthorn$^{6}$,\and
    Satoru~Iguchi$^{7,8}$,
    Themiya~Nanayakkara$^{6,9}$,
    Karl~Glazebrook$^{9}$\and
    \small$^{1}$Kavli Institute for the Physics and Mathematics of the Universe (WPI), The University of Tokyo, Kashiwa, Japan.\and
    \small$^{2}$Astronomical Institute, Tohoku University, Sendai, Japan.\and
    \small$^{3}$Research School of Astronomy and Astrophysics, Australian National University, Canberra, Australia.\and
    \small$^{4}$Department of Astronomy, School of Physics, Peking University, Beijing, People's Republic of China.\and
    \small$^{5}$Kavli Institute for Astronomy and Astrophysics, Peking University, Beijing, People's Republic of China.\and
    \small$^{6}$Sydney Institute for Astronomy, The University of Sydney, Sydney, Australia.\and
    \small$^{7}$Graduate Institute for Advanced Studies (SOKENDAI), Mitaka, Japan.\and
    \small$^{8}$National Astronomical Observatory of Japan, National Institutes of Natural Sciences, Mitaka, Japan.\and
    \small$^{9}$Centre for Astrophysics and Supercomputing, Swinburne University of Technology, Melbourne, Australia.\and
    \small$^\ast$Corresponding author. Email: tsukuitk23@gmail.com
}

\begin{document} 

% Insert the title and author list
\maketitle

% Abstract, in bold
% There are strict length limits, and not all formats have abstracts.
% Consult the journal instructions to authors for details.
% Do not cite any references in the abstract.
\begin{abstract} \bfseries \boldmath
Disk galaxies like the Milky Way are thought to evolve through internal dynamical processes: the stellar disk forms a bar, the bar drives gas inflow that builds a nuclear stellar disk, and the bar vertically thickens into an X-shaped bulge. Although this evolution is thought to be slow, completing only at late cosmic times, its timing remains poorly constrained. We report James Webb Space Telescope imaging of a galaxy at redshift 0.92 (7.6 billion years ago) that already hosts an X-shaped bulge, a nuclear stellar disk, and an extended stellar disk, with geometry and inferred bar size indistinguishable from those of present-day barred galaxies. The X-shaped bulge marks the completion of the major phase of bar-driven evolution when the Universe was less than half its current age.% and requires the stellar disk to have settled much earlier.
\end{abstract}

% The first paragraph of any Science paper does NOT have a heading
% Nor is it indented

\noindent
Galaxies form as highly disturbed gas settles into ordered rotating disks \cite{genzelRapid2006}. Once such a disk forms, it is thought to evolve through gravitational disk instabilities \cite{kormendySecular2004}: the stellar disk forms a bar \cite{ostrikerNumerical1973}, the bar drives gas inflow \cite{athanassoulaExistence1992, wadaRapid1992} and builds a nuclear stellar disk \cite{gadottiMUSE2015, babaAge2020}. Once a bar grows strong enough, it becomes unstable: its inner part thickens through violent buckling instability or vertical heating \cite{combesBox1990, rahaDynamical1991} and becomes an X-shaped bulge, often seen in nearby edge-on disk galaxies \cite{savchenkoMeasuring2017}, including the Milky Way \cite{nessXshaped2016}. This evolution is thought to be slow and to complete at relatively late cosmic epochs \cite{kormendySecular2004, shethHot2012}, while its onset remains poorly constrained \cite{krukRevealing2019, lopezFormation2025}. James Webb Space Telescope (JWST) has recently found massive stellar disks as early as $z\sim6$ \cite{wangGiant2025, jainGranddesign2025, xiaoPANORAMIC2025} and barred galaxies up to $z\sim3$ (2.1 Gyr after the Big Bang) \cite{leconteJWST2024, guoFirst2023, costantinMilky2023, geronGalaxy2025}. Independent evidence for dynamically mature bars at high redshift comes from JWST observations of face-on barred galaxies at $z\sim1.5$, which reveal flattened stellar bar surface-brightness profiles \cite{kalitaGalactic2026}, commonly associated with dynamically evolved bars \cite{BureauKband2006, andersonSecular2022} and nuclear stellar disks \cite{leconteNuclear2026}. Complementary stellar-age measurements imply bar formation at $\sim 8 \pm 1$ Gyr ago in the Milky Way \cite{sandersEpoch2024} and over a wider range of 1–13 Gyr in nearby galaxies \cite{desa-freitasBar2025}. Distant bars have so far been studied almost exclusively in face-on orientations, where bars are most easily detected; an edge-on view instead uniquely reveals the vertical evolution of the bar.

\begin{figure}
    \centering
    \includegraphics[width=0.325\linewidth]{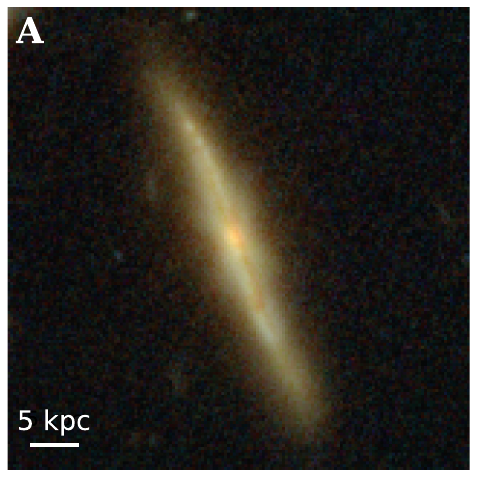}
    \includegraphics[width=0.325\linewidth]{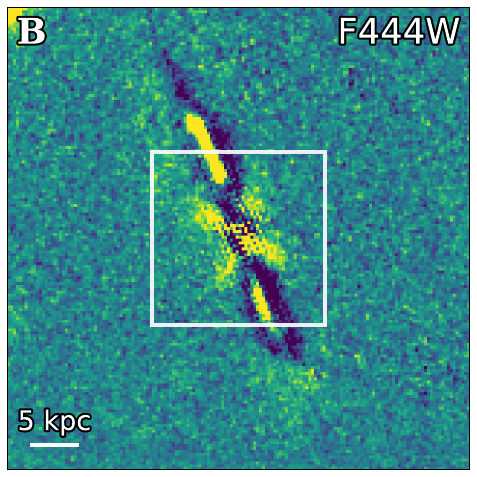}
    \includegraphics[width=0.325\linewidth]{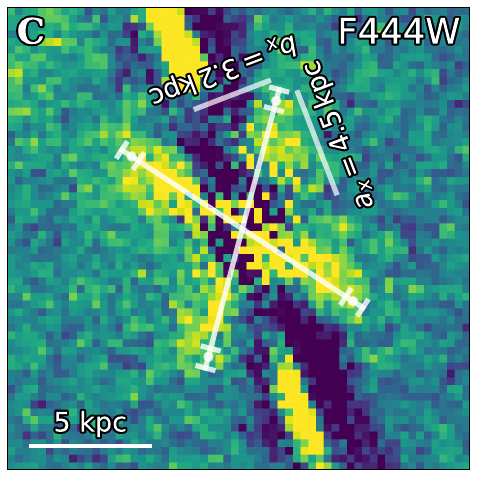}
    \caption{\textbf{Stellar structure and X-shaped bulge of UDS 12999.} (\textbf{A}) A color composite image of UDS 12999 constructed from JWST wide-band filters F090W to F444W. The earliest known X-shaped bulge extends from the galaxy center. (\textbf{B}) The feature-enhanced F444W image obtained after subtracting the best-fit disk and S\'ersic models. (\textbf{C}) A zoomed-in view of the region marked by the white square in panel (B), with the measured structural parameters of the X-shaped structure overlaid.}
    \label{fig:Fig1}
\end{figure}

\subsection*{Discovery of an X-shaped bulge at z = 0.92}
We identify a clear X-shaped structure in the edge-on galaxy UDS 12999, 7.6 Gyr ago, from the James Webb Space Telescope Near Infrared Camera (JWST/NIRCam) images taken as part of the Public Release Imaging for Extragalactic Research (PRIMER) survey \cite{dunlopPRIMER2021}. Figure~\ref{fig:Fig1}A shows a color composite image of the galaxy, in which the X-shaped structure is visible, defined by four rays extending from the center of the edge-on disk. The feature is more evident in the substructure-enhanced images (Fig.~\ref{fig:Fig1}B-C), which were made by subtracting the overall best-fit galactic components of the extended disk and the compact S\'ersic component from the F444W image \cite{methods}. The X-shaped structure is intrinsic, not a point spread function (PSF) artifact \cite{methods}. We measure the source redshift, $z=0.92331\pm0.00003$, spectroscopically with the JWST Near Infrared Spectrograph (JWST/NIRSpec) \cite{methods}, making UDS 12999 the most distant galaxy currently known to host such a feature. The eight-band JWST/NIRCam images ($0.9-4.4$~$\mu$m) of UDS 12999 are well described by three distinct components: an extended disk, a central compact component, and an X-shaped structure \cite{methods}. We measure their structural and spectral energy distribution (SED) properties (Table~\ref{tab:tab1}) by simultaneous multi-wavelength structural modeling, masking the dust lane and accounting for the PSF \cite{methods}. 

\subsection*{Structural resemblance to present-day barred galaxies}
\begin{figure}
    \centering
    \includegraphics[width=\linewidth]{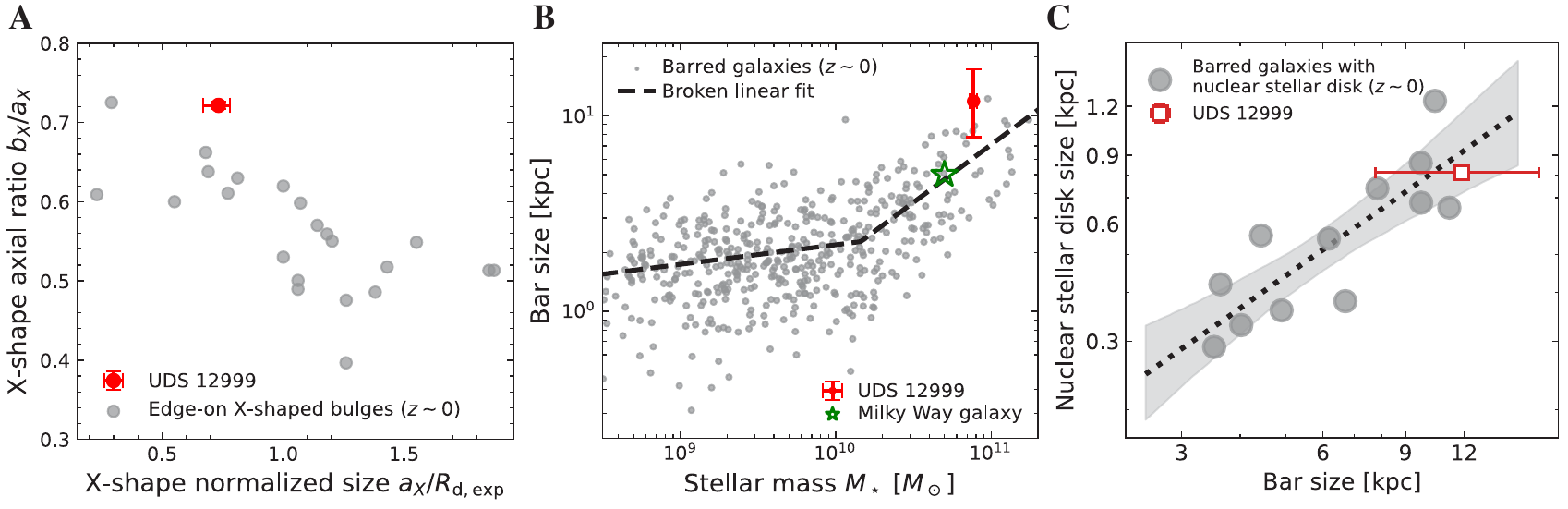}
    \caption{\textbf{Structural comparison between UDS 12999 and present-day galaxies.} (\textbf{A}) UDS 12999 is compared with nearby edge-on galaxies in the plane of the X-shaped bulge minor-to-major axis ratio, $b_X/a_X$, versus the major-axis length $a_X$, normalized by the exponential disk scale length $R_{\rm d, exp}$. Nearby measurements are from \cite{savchenkoMeasuring2017} and were derived using a method similar to ours \cite{methods}. (\textbf{B}) Estimated bar sizes versus stellar mass for UDS 12999, the Milky Way \cite{weggStructure2015, bland-hawthornGalaxy2016}, and nearby barred galaxies \cite{erwinWhat2019}. The dashed line shows the broken linear fit \cite{erwinWhat2019}. (\textbf{C}) Estimated bar sizes versus nuclear stellar disk size. The measured S\'ersic-component sizes \cite{saloSpitzer2015} are shown as nuclear stellar disk sizes for $z=0$ galaxies in which the S\'ersic component is kinematically confirmed to be a nuclear stellar disk \cite{gadottiKinematic2020}. The dashed line shows the orthogonal linear regression fit to the $z=0$ sample, and the shaded region indicates the 1$\sigma$ bootstrap uncertainty.}
    \label{fig:Fig2}
\end{figure}

The geometry and size of the X-shaped structure are consistent with the sequence formed by local edge-on galaxies with X-shaped bulges \cite{savchenkoMeasuring2017} in the plane of minor-to-major axis ratio $b_X/a_X$ versus the major-axis extent $a_X$, normalized by the disk scale length $R_{\rm d, exp}$ (Fig.~\ref{fig:Fig2}A and Table~\ref{tab:tab1}). The relation and its scatter reflect both intrinsic structure and projection effects: the intrinsic axial ratio $b_X/a_X$ may evolve as bars grow longer \cite{ciamburDouble2021}, while the observed axial ratio depends on the X-shaped bulge's azimuthal viewing angle relative to the side-on orientation \cite{savchenkoMeasuring2017}. Given this similarity to local X-shaped bulges, we estimate a bar semi-major axis (hereafter, bar size) of $a_\textrm{bar}=11.8^{+5.5}_{-4.0}$ kpc using the ratio $a_X/a_\text{bar}=0.26-0.58$ (mean $0.38$) observed in local barred galaxies \cite{erwinPeanuts2013, laurikainenBarlenses2017}. This range reflects the variation in viewing angle and thus encompasses the projection uncertainty. UDS 12999 follows the local relation between bar size and stellar mass (Fig.~\ref{fig:Fig2}B), suggesting that its bar and X-shaped structure formed through the same mass-dependent disk dynamical processes as in nearby barred galaxies \cite{erwinWhat2019}. 

The central compact component has a half-light radius $R_S=812\pm9$ pc, ellipticity $e_S=0.72\pm0.01$, and S\'ersic index $n_S=0.84^{+0.05}_{-0.05}$, close to the exponential profile ($n=1$) typical of disks. Its high ellipticity likewise indicates rotational support. Such compact, flat, exponential components in barred galaxies are typically associated with nuclear stellar disks \cite{gadottiKinematic2020}, which form from bar-driven gas inflow shortly after bar formation \cite{gadottiMUSE2015, babaAge2020}. The measured compact disk size and the estimated bar size follow the local relation for barred galaxies with kinematically confirmed nuclear stellar disks (Fig.~\ref{fig:Fig2}C). We therefore identify this component as a nuclear stellar disk. Together, these results (Fig.~\ref{fig:Fig2}) show that the key structural products of bar-driven secular evolution -- a bar, an X-shaped bulge, and a nuclear stellar disk -- are already in place in UDS 12999 only 6.2 Gyr after the Big Bang. These structures are indistinguishable from those in local galaxies.

\subsection*{Star formation and mass assembly history}

\begin{figure}
    \centering
    \includegraphics[width=1\linewidth]{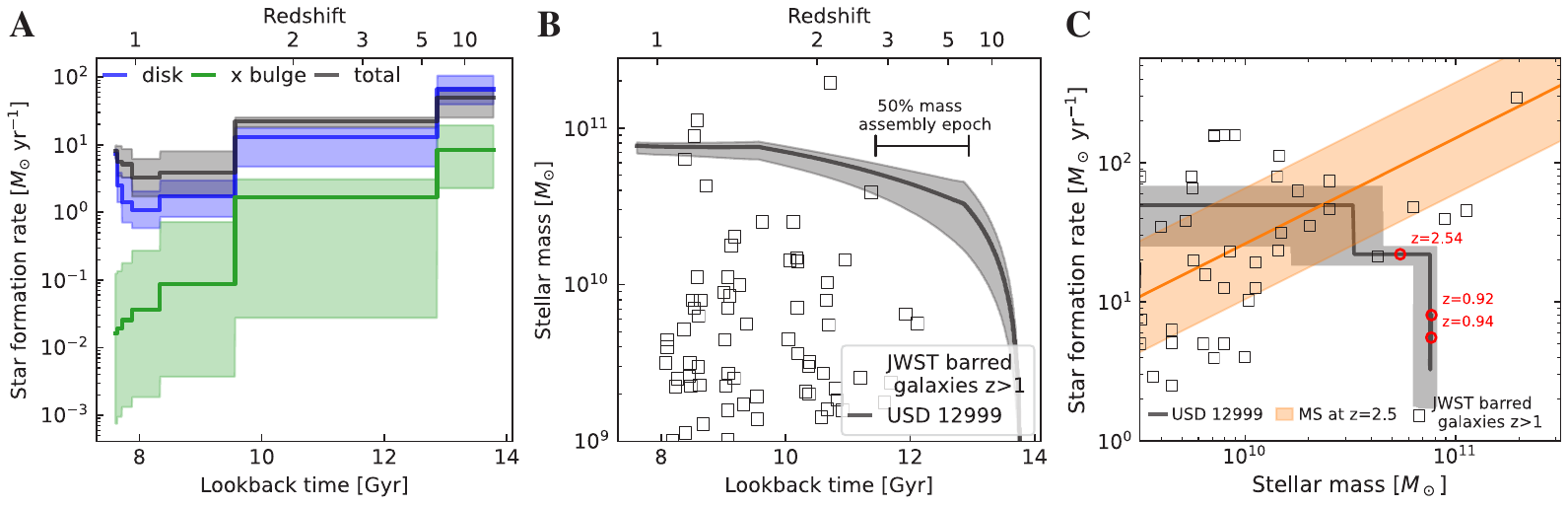}
    \caption{\textbf{Star formation and stellar mass assembly history of UDS 12999.} (\textbf{A}) The SFH of UDS 12999 (black) and its decomposed disk (blue) and X-shaped bulge (green) components. (\textbf{B}) The stellar mass growth history of the galaxy. (\textbf{C}) Its evolutionary track in the stellar mass--star formation rate plane. The main sequence relation for typical star-forming galaxies at $z=2.5$ from ref. \cite{speagleHighly2014} is shown in panel (C) for comparison. These quantities are derived from the spectral energy distribution modeling (SED) of HST and JWST photometric data using \textsc{Prospector} \cite{johnsonStellar2021} with 1$\sigma$ statistical uncertainties shown by shaded regions. Barred galaxies identified in JWST imaging at $z>1$ \cite{geronGalaxy2025, coxCeers2025, costantinMilky2023} are overplotted in panels (B) and (C).}
    \label{fig:Fig3}
\end{figure}

Figure~\ref{fig:Fig3} traces the assembly history of UDS 12999: (\textbf{A}) the star formation history (SFH), (\textbf{B}) the resulting stellar mass growth, and (\textbf{C}) the trajectory in the stellar mass - star formation rate plane. All quantities are inferred from stellar population synthesis modeling constrained by photometric measurements from the Hubble Space Telescope Advanced Camera for Surveys (HST/ACS) and Wide Field Camera 3 (HST/WFC3) and the JWST/NIRCam dataset using the \textsc{Prospector} code \cite{johnsonStellar2021, methods}. The total SFH is constrained by the aperture photometry that encloses the entire galaxy (Table~\ref{tab:tab2}). Separate estimates are obtained for the extended disk and X-shaped components using the photometry of each structurally decomposed component \cite{methods}.

The SFH shows an early phase of intense star formation with an average rate of $50^{+19}_{-24}$ $M_\odot \mathrm{yr}^{-1}$ more than 13 Gyr ago, followed by a prolonged period of low star formation ($<10\,M_\odot \mathrm{yr}^{-1}$ since 9.5 Gyr ago; Fig.~\ref{fig:Fig3}A). Accordingly, the galaxy formed most of its stellar mass early, with half of the mass already in place 11.4–13.0 Gyr ago ($z=2.7-6.9$; Fig.~\ref{fig:Fig3}B), and reached a total stellar mass of $7.7^{+0.4}_{-0.8} \times 10^{10}M_\odot$ at 7.6 Gyr ago, already about 1-2 times that of the Milky Way \cite{bland-hawthornGalaxy2016}. After this main stellar assembly phase, the galaxy evolved with suppressed star formation, at a rate below that of typical star-forming galaxies at each epoch. It lies below the star-forming main sequence (SFMS) from $z=2.5$ to the observed epoch ($z=0.92$; Fig.~\ref{fig:Fig3}C), whereas barred galaxies at similar redshifts ($z=1-3$) identified by JWST \cite{geronGalaxy2025, costantinMilky2023} lie on the main sequence. These results suggest that UDS 12999 grew rapidly early on, placing it in the upper mass range of JWST-detected barred galaxies at $z=1–3$. It then evolved through an extended phase of suppressed star formation. Given that mergers at $z>1$ are typically gas-rich and trigger star formation \cite{linRedshift2008}, the absence of such enhancement suggests that the galaxy experienced a quiet merger history over the past $\sim$3.5 Gyr ($z=2.5-0.92$). 

Figure~\ref{fig:Fig3}A compares the SFHs of the disk and X-shaped bulge. Once normalized by their peak star formation rate in the oldest age bin, they are consistent with each other until 270 million years ago, but diverge more recently; star formation ceased in the X-shaped bulge, while the disk continued to form stars. This is consistent with a picture in which the X-shaped structure is made of stars lifted from the disk plane into stable resonant orbits supported by a rotating bar structure \cite{sellwoodThree2020, babaAge2022}. The process that lifts stars into the X-shaped bulge may have become less efficient $\sim$ 270 million years ago.

\subsection*{A tight time budget for bar-driven evolution}

The presence of an X-shaped bulge in UDS 12999 places a strong constraint on how rapidly the galaxy assembled its dynamically settled disk. An X-shaped bulge marks the endpoint of a sequential bar-driven evolution process: the disk must first settle, a bar must form, and the bar must develop an X-shape \cite{combesBox1990, rahaDynamical1991, sellwoodThree2020}. We observe UDS 12999 at 7.6 Gyr ago, with roughly half of its stellar mass assembled 11.4-13 Gyr ago, leaving only 3.8–5.4 Gyr for the galaxy to develop an X-shaped bulge after its main assembly. Therefore,
\begin{equation}
\tau_{\rm{disk}}+\tau_{\rm{bar}}+\tau_{\rm{X}}< \textrm{3.8-5.4 Gyr},
\end{equation}
where $\tau_{\rm{disk}}$ is the time required for the disk to settle after half of the stellar mass assembled, $\tau_{\rm{bar}}$ is the time required for bar formation after the disk forms, and $\tau_{\rm{X}}$ is the time needed for the bar to develop an X-shaped bulge. Simulations of disk galaxies typically require $\tau_{\rm bar}\sim2$ Gyr for bar formation \cite{fujiiDynamics2018, bland-hawthornRapid2023, asanoExponential2026}, although the timescale depends strongly on disk properties and baryon fraction \cite{fujiiDynamics2018, bland-hawthornRapid2023, seoEffects2019, fragkoudiBar2025}. The bar then requires another $\tau_{\rm{X}}\sim1$–2 Gyr to develop a prominent X-shape through a buckling event \cite{asanoExponential2026, lopezFormation2025}. However, some simulations suggest that X-shaped structures can also emerge without a strong buckling phase, potentially shortening the X-shaped bulge formation time \cite{babaAge2022, seoEffects2019}. Even allowing for these variations, the available time budget remains tight, implying that a dynamically settled disk formed soon after the galaxy assembled half of its stellar mass ($z=2.7-6.9$). This provides independent dynamical evidence for early disk systems. 

In general, massive galaxies complete star formation earlier \cite{behrooziUniverseMachine2019}, form cold stellar disks earlier \cite{tsukuiEmergence2025}, and are more likely to host strong bars \cite{leconteJWST2024}. Consistent with this picture, the early rapid stellar mass assembly of UDS 12999 may have favored early disk settling and bar formation. Strong gas dissipation during rapid galaxy assembly leads to baryon dominance in the inner disk \cite{genzelStrongly2017}, which may further shorten the bar-formation timescale $\tau_{\rm bar}$ \cite{fujiiDynamics2018, bland-hawthornRapid2023, fragkoudiBar2025}.

The clear X-shaped bulge indicates that the disk already had a fraction of stars on cold orbits when buckling occurred, a necessary condition suggested by simulations \cite{ghoshBars2024}. Given its stellar mass and observed epoch, UDS 12999 is expected to lie at the transition between thick-disk formation and subsequent thin-disk growth \cite{tsukuiEmergence2025, yuThick2026}, consistent with its current suppressed star formation after the burst. The cold stars need not belong to a separately formed thin disk: in the emerging picture of disk evolution, the stars now in the thick disk formed close to the mid-plane on cold orbits and thickened quickly \cite{tsukuiEmergence2025, bland-hawthornTurbulent2025, yuThick2026}. An early thick disk can therefore still retain a substantial population of such cold stars. The coexistence of hot and cold stellar disks is suggested to further reduce the bar-buckling time $\tau_{\rm{X}}$ from about 2.5 to 1 Gyr \cite{ghoshBars2024}.

Earlier cosmological simulations predicted that X-shaped bulges form only at late times \cite{debuhrStellar2012} ($z=0.5-0.7$). More recent work, however, shows that X-shaped bulges can emerge through buckling as early as a lookback time of $\sim$8 Gyr but remain vertically asymmetric at first, becoming symmetric, as observed in UDS 12999, only at lookback times $\lesssim$7.6 Gyr, as shown in the snapshot sequence in figure~12 of \cite{lopezFormation2025}. The symmetric X-shaped bulge in UDS 12999 is consistent with the earliest such structures formed in simulations. However, small differences in the initial conditions of simulations can lead to substantial variations in the X-shaped bulge formation time \cite{asanoExponential2026}, limiting direct comparisons with a small number of simulated galaxies. 

\subsection*{A dynamically supported bar at early times}

At high redshift, stellar bars are now routinely identified by their elongated morphology in stellar images of face-on galaxies \cite{leconteJWST2024, guoFirst2023, geronGalaxy2025}. Morphology alone cannot establish that such structures are long-lived stellar bars, self-supported by stellar orbits, rather than transient features. Seen edge-on, UDS 12999 uniquely captures the earliest known X-shaped bulge, which is the thickened inner part of a stellar bar and a robust outcome of its dynamical evolution, established in N-body simulations \cite{combesBox1990, rahaDynamical1991, asanoExponential2026}. Its detection supports the broader interpretation that elongated stellar morphologies at high redshift can trace genuine, long-lived stellar bars. Overall, the structural and chronological properties of UDS 12999 fit and strengthen an emerging picture of early bar-driven secular evolution.

\begin{table}
\centering
\caption{\textbf{Measured structural and stellar population properties of UDS 12999.} Errors denote 1$\sigma$ statistical uncertainties. The systematic uncertainty in $a_{\rm bar}$ reflects the observed range of $a_X/a_{\rm bar}$ in local barred galaxies, effectively including the uncertain bar viewing angle \cite{erwinPeanuts2013, laurikainenBarlenses2017}. $R_{\rm d, exp}$ is derived from a pure exponential disk fit, whereas $R_{\rm d, eff}$ is measured from the fiducial structural fit, allowing deviations from an exponential profile \cite{methods}.} 
\label{tab:tab1}
\begin{tabular}{ll}
\hline
\hline
Property & Value \\
\hline
Galaxy name & UDS 12999 \\
Right Ascension (RA) & 34.3119 deg \\
Declination (Dec) & -5.23517 deg \\
Redshift, $z$ & $0.92331 \pm 0.00003$ \\
Stellar mass, $M_\star$ & $7.7^{+0.4}_{-0.8} \times 10^{10}M_\odot$ \\
Star-formation rate & $8.0^{+1.7}_{-1.7} M_\odot \rm yr^{-1}$ \\
\hline
Extended disk structure\\
\hline
Thickness, $z_0$ & $578^{+1}_{-2}$ pc \\
Half-light radius, $R_{\rm d, eff}$ & $11.28^{+0.01}_{-0.01}$ kpc \\ 
Exponential scale radius, $R_{\rm d, exp}$ & $6.13^{+0.02}_{-0.01}$ kpc \\
\hline
X-shaped/bar structure\\
\hline
Semi-major axis, $a_X$ & $4.5^{+0.4}_{-0.3}$ kpc \\
Axis ratio, $b_X/a_X$ & $0.72\pm0.01$ \\
Estimated bar size, $a_\textrm{bar}$ & $11.8^{+1.1}_{-0.8}$(stat)$^{+5.5}_{-4.0}$(sys) kpc\\
\hline
Nuclear stellar disk (S\'ersic component)\\
\hline
Half-light radius, $R_\textrm{s}$ & $812^{+9}_{-9}$ pc\\
S\'ersic index, $n_\textrm{s}$ & $0.84^{+0.05}_{-0.05}$\\
Ellipticity, $e_\textrm{s}$ & $0.72\pm{0.01}$\\
\hline
\end{tabular}
\end{table}
%%%%%%%%%%%%%%%% REFERENCES %%%%%%%%%%%%%%%

\clearpage % Clear all remaining figures and tables then start a new page

% The list of references goes after the main text and before the acknowledgements
% When preparing an initial submission, we recommend you use BibTeX, like this:
%
\bibliography{Tsukui26} % for a file named science_template.bib
\bibliographystyle{sciencemag}

% After the paper has completed peer review and been revised ready for acceptance,
% you should comment out the lines above and copy-paste the contents of your .bbl
% file here instead. This will help ensure that our conversion software works correctly.
% Remember to re-run BibTeX first - check the timestamp!
%
% Example of the first three entries copy-pasted from science_template.bbl:
%
%\begin{thebibliography}{1}
%
%\bibitem{example}
%A.~N. {Author}, An example reference. \emph{Journal of Improbable Research}
%  \textbf{1}, 67 (2020).
%
%\bibitem{example2}
%F.~M. {Surname}, S.~{Author}, A second example. \emph{Interesting Research
%  Letters} \textbf{32}, 897 (2019).
%
%\bibitem{example_preprint}
%P.~{One}, P.~{Two}, P.~{Three}, {An unpublished preprint}. \emph{preprint}
%  (2021), arXiv:2101.12345.
%
%\end{thebibliography}

%%%%%%%%%%%%%%%% ACKNOWLEDGEMENTS %%%%%%%%%%%%%%%

\section*{Acknowledgments}
T.T. thanks Ken Freeman, Yoshihisa Asada, Junichi Baba, Shogo Nishiyama, John Silverman, Thorsten Tepper Garcia, Daisuke Taniguchi, and Martin Bureau for helpful discussions. This work is based on observations made with the NASA/ESA/CSA James Webb Space Telescope. This work is supported by the High-performance Computing Platform of Peking University. The data products presented herein were retrieved from the Dawn JWST Archive (DJA). DJA is an initiative of the Cosmic Dawn Center (DAWN), which is funded by the Danish National Research Foundation under grant DNRF140. 
\paragraph*{Funding:}
 This work was supported in part by Japan Foundation for Promotion of Astronomy and the Hayakawa Fund of the Astronomical Society of Japan, which provided travel support to T.T. T.T. is supported by the JSPS Grant-in-Aid for Research Activity Start-up (25K23392) and the JSPS Core-to-Core Program (JPJSCCA20210003). T.N. is supported by Australian Research Council Discovery Project DP230103161.
\paragraph*{Author contributions:}
T.T. identified the X-shaped bulge in the galaxy and led the data analysis and manuscript preparation. Z.G. contributed to the stellar population analysis. T.N. and K.G. helped T.T. analyze the JWST/NIRSpec data. All authors contributed to the discussion, interpretation of the results, and the writing of the paper.
\paragraph*{Competing interests:}
The authors declare that they have no competing interests.
\paragraph*{Data, code, and materials availability:}
All JWST/HST data used in this study are publicly available from the DAWN JWST Archive (DJA; \url{https://dawn-cph.github.io/dja/index.html}) and the Mikulski Archive for Space Telescopes (MAST; \url{http://archive.stsci.edu}). All code used in this study is publicly available. The structural modeling tool developed in this study is publicly available \cite{tsukuiTakafumi2912026}.

%%%%%%%%%%%%%%%% SUPPLEMENT LIST %%%%%%%%%%%%%%%
\subsection*{Supplementary materials}
Materials and Methods\\
Figs. S1 to S8\\
Tables S1 to S3\\
References \textit{(57-\arabic{enumiv})}\\ % automatically fills out the last reference number
% (filling out the other numbers automatically is possible but fiddly and liable to break)
%%%%%%%%%%%%%%%% END OF MAIN TEXT %%%%%%%%%%%%%%%

\newpage

%%%%%%%%%%%%%%%% START OF SUPPLEMENT %%%%%%%%%%%%%%%

% Figures, tables, equations and pages in the supplement are numbered S1, S2 etc.
\renewcommand{\thefigure}{S\arabic{figure}}
\renewcommand{\thetable}{S\arabic{table}}
\renewcommand{\theequation}{S\arabic{equation}}
\renewcommand{\thepage}{S\arabic{page}}
\setcounter{figure}{0}
\setcounter{table}{0}
\setcounter{equation}{0}
\setcounter{page}{1} % not 0 as \newpage already started a supplementary page
% References continue the numbering from the main text.

%%%%%%%%%%%%%%%% SUPPLEMENT TITLE PAGE %%%%%%%%%%%%%%%

\begin{center}
\section*{Supplementary Materials for\\ \scititle}

Takafumi~Tsukui$^{\ast}$,
Zeyu~Gao,
Emily~Wisnioski,
Joss~Bland-Hawthorn,\\
Satoru~Iguchi,
Themiya~Nanayakkara,
Karl~Glazebrook\\
\small$^\ast$Corresponding author. Email: tsukuitk23@gmail.com
% Author list for the supplement
% Indicate the corresponding authors, but do NOT include institutions here
% It would be nice if the template auto-generated this, but doing so is complicated...
\end{center}
% Fill out the numbers for each type of supplementary material,
% and delete any lines that aren't applicable.
% These are just example numbers that don't match the rest of this template.
\subsubsection*{This PDF file includes:}
Materials and Methods\\
Figures S1 to S8\\
Tables S1 to S3\\
\newpage

%%%%%%%%%%%%%%%% MATERIALS AND METHODS %%%%%%%%%%%%%%%

\section*{Materials and Methods}
\subsection*{Cosmology and initial mass function}
Throughout the paper, we adopted a flat $\Lambda$CDM cosmology with a present-day Hubble constant of $H_0=67.7$ km s\textsuperscript{-1} Mpc\textsuperscript{-1} and matter density in units of the critical density $\Omega_\mathrm{m}=0.310$ \cite{planckcollaborationPlanck2020}. The angular diameter distance of the source is $D_{\mathrm{A}}$=1664.14 Mpc for a source at redshift $z=0.92331$ \cite{wrightCosmology2006}. We assumed a Chabrier initial mass function \cite{chabrierGalactic2003}. 

\subsection*{The galaxy UDS 12999}
The galaxy UDS 12999 (CANDELS UDS F160W J021714.84-051406.5; RA: 34.3119 deg, Dec: -5.23517 deg) is located in the UKIRT Infrared Deep Sky Survey Ultra-Deep Survey field (UKIDSS UDS) \cite{lawrenceUKIRT2007} and was observed by the Hubble Space Telescope as part of the Cosmic Assembly Near-infrared Deep Extragalactic Legacy Survey (CANDELS) Ultra Deep Survey \cite{groginCANDELS2011, koekemoerCANDELS2011}. We adopt the name UDS 12999 from the 3D-HST photometric catalog \cite{skelton3DHST2014}. Previous photometric and grism measurements placed the galaxy at $z=0.84-1.04$ \cite{momcheva3DHST2016, straatmanFourStar2016, merlinASTRODEEPJWST2024}. Below, we measure a spectroscopic redshift of $z = 0.92331 \pm 0.00003$ using JWST/NIRSpec data.

\subsection*{JWST/HST imaging}
We used publicly released HST/JWST mosaic images from the DAWN \textit{JWST} Archive (DJA), which were reduced using the \textsc{grizli} pipeline \cite{brammerGrizli2023, valentinoAtlas2023}. All images were sampled to a common pixel scale of 0.04 arcsec per pixel. The JWST imaging came from the Public Release Imaging for Extragalactic Research (PRIMER) survey \cite{dunlopPRIMER2021}, whereas the HST imaging combined data from multiple programs (IDs 12064, 12099, 12461, 15862, 12328, 15363, 13002, 16872, 10876, and 13792). We made the color composite image (Fig.~\ref{fig:Fig1}A) by combining all JWST wide-band filters using the \textsc{Trilogy} package \cite{coeCLASH2012} with default parameters.

We measured the total flux within an elliptical aperture that conservatively encloses the entire galaxy using the \textsc{Photutils} package \cite{larrybradleyAstropy2019}. The aperture was defined from the source segmentation mask produced using \textsc{SExtractor} on stacked images of the long-wavelength filters (F277W, F356W, and F444W) \cite{valentinoAtlas2023}. We expanded the segmentation mask by 3 pixels ($=0.12"$), corresponding to roughly the FWHM of the most extended PSF (F444W). We estimated the blank-sky level and subtracted the corresponding sky contribution from the total flux within the aperture, while propagating the uncertainties. Table~\ref{tab:tab2} summarizes the available HST/JWST bands and measured photometry. For structural modeling of the JWST images, we used the empirical PSFs constructed from images of unsaturated stars by ref. \cite{geninDAWN2025}.

\begin{table}
\centering
\caption{\textbf{Integrated photometry of UDS 12999.} Photometry is measured within an elliptical aperture enclosing the galaxy. The reference wavelengths are taken from the SVO Filter Profile Service \cite{rodrigoSVO2020}.}
\label{tab:tab2}
\begin{tabular}{lcc}
\hline
\hline
Band & $\lambda_{\rm ref}$ ($\mu$m) & Flux ($\mu$Jy) \\
\hline
\multicolumn{3}{l}{\textit{JWST / NIRCam}} \\
\hline
F090W & 0.90 & $3.551 \pm 0.107$ \\
F115W & 1.15 & $6.900 \pm 0.100$ \\
F150W & 1.50 & $10.367 \pm 0.084$ \\
F200W & 2.00 & $16.897 \pm 0.069$ \\
F277W & 2.76 & $24.562 \pm 0.049$ \\
F356W & 3.57 & $28.035 \pm 0.049$ \\
F410M & 4.08 & $26.730 \pm 0.091$ \\
F444W & 4.40 & $23.231 \pm 0.070$ \\

\hline
\multicolumn{3}{l}{\textit{HST / ACS}} \\
\hline
F435W & 0.43 & $0.355 \pm 0.145$ \\
F606W & 0.59 & $0.460 \pm 0.065$ \\
F814W & 0.80 & $2.382 \pm 0.074$ \\

\hline
\multicolumn{3}{l}{\textit{HST / WFC3-IR}} \\
\hline
F125W & 1.25 & $7.323 \pm 0.118$ \\
F140W & 1.39 & $8.874 \pm 0.213$ \\
F160W & 1.54 & $10.793 \pm 0.105$ \\
\hline
\end{tabular}
\end{table}

\subsection*{JWST NIRSpec spectroscopy and redshift}

\begin{figure}
    \centering
    \includegraphics[width=1\linewidth]{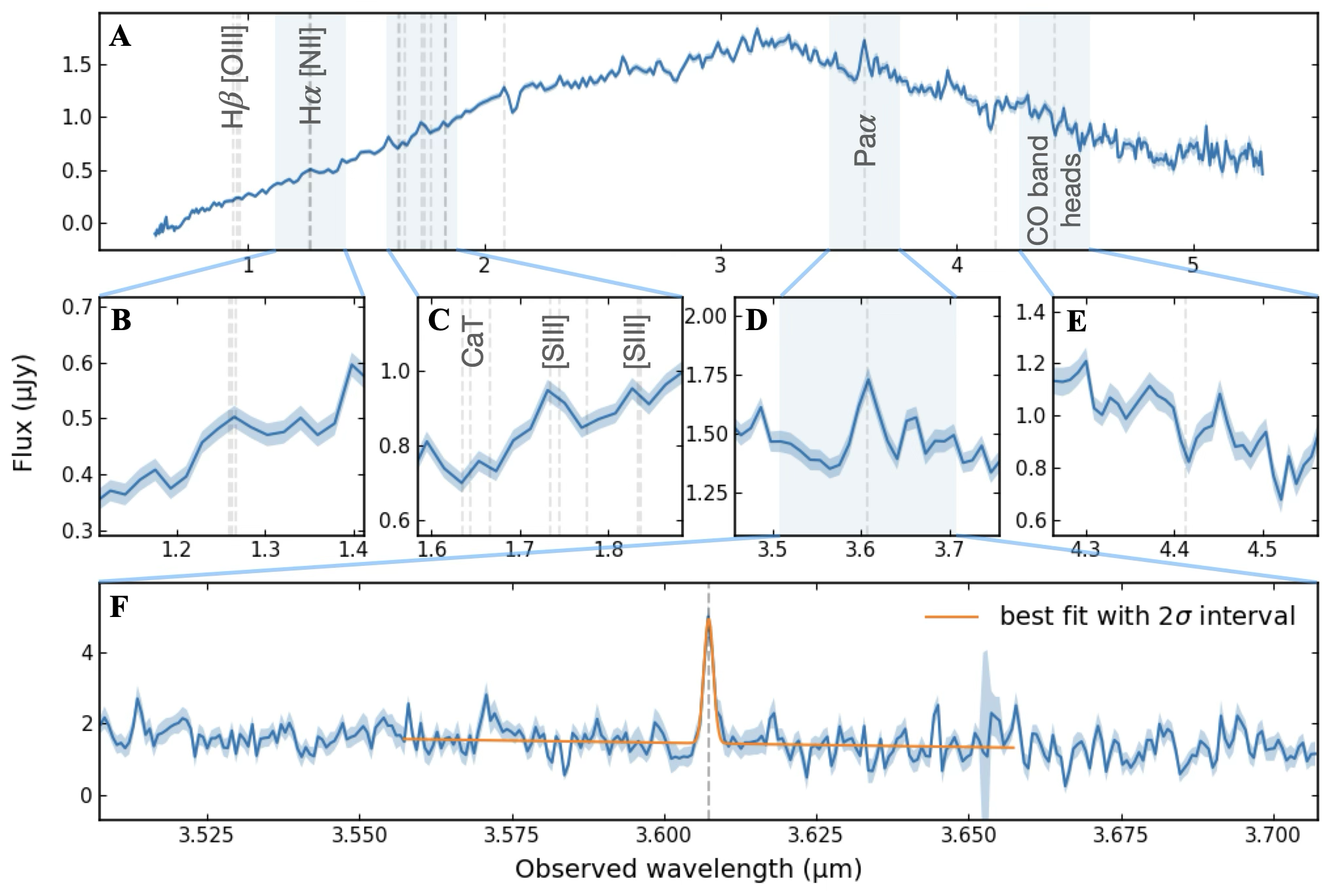}
    \caption{\textbf{JWST NIRSpec spectra for UDS 12999.} (\textbf{A}) Low resolution PRISM spectrum with major emission lines marked by vertical dashed lines. (\textbf{B}-\textbf{E}) Zoomed-in views around the four major lines highlighted by shaded regions in panel (A). (\textbf{F}) The high-resolution spectrum (G395H-F290LP) around the Pa$\alpha$ emission line. The orange curve shows the best-fit model, consisting of a Gaussian emission line profile and a linear continuum, from which we determine the redshift.}
    \label{fig:Fig4}
\end{figure}

UDS 12999 was observed with JWST NIRSpec as part of the Guaranteed Time Observations program NIRSpec WIDE MOS Survey (ID 1215; PI=Nora Luetzgendorf). A slitlet dedicated to background measurement was placed on the galaxy, using the PRISM-CLEAR, G235H-F170LP, and G395H-F290LP disperser-filter combinations. We reduced the data using the publicly available pipeline \textsc{jwst v1.18.0} with the JWST Calibration Reference Data System (CRDS) context file \texttt{jwst\_1364.pmap} provided by STScI. Because the galaxy extends across the slitlet, the standard nod-based background subtraction could not be applied. Instead, we constructed a master background from nearby slitlets within the same quadrant of the micro-shutter assembly and subtracted it from each exposure. We then masked bad pixels and combined the exposures into rectified, wavelength-calibrated and flux-calibrated 2D spectral images. We extracted the 1D spectrum of the target by integrating over the high signal-to-noise spatial region.

The reduced PRISM spectra are shown in Fig.~\ref{fig:Fig4}A-E, where several strong emission and absorption features are detected, most notably the calcium triplet (CaT) absorption lines, the Paschen-$\alpha$ (Pa$\alpha$) emission line, and the $\sim2.29$ $\mu$m carbon monoxide (CO) bandheads. The high-resolution spectrum with G395H-F290LP covers the Pa$\alpha$ line and is shown in Fig.~\ref{fig:Fig4}F, allowing us to determine the redshift of the galaxy. By fitting the detected Pa$\alpha$ line (observed-frame 0.1 $\mu$m window around the line center) with a five-parameter model consisting of a Gaussian emission-line profile (mean, width, and amplitude) and a linear continuum (slope and intercept) using the \textsc{emcee} package \cite{foreman-mackeyEmcee2013}, we find a galaxy redshift of $z=0.92331\pm0.00003$. The observation covered only the outer region of the disk, so we did not use this spectrum to derive the global galaxy properties.

\subsection*{Structural modeling of the galaxy}

To measure the structural properties and spectral energy distributions (SEDs) of the main stellar subcomponents in UDS 12999, we performed a multi-component structural decomposition of the multi-band photometric images. The model is intended to capture the structure and color distribution of the galaxy with a minimal number of components and parameters. We find that three components -- standard disk and S\'ersic components used in edge-on disk modeling \cite{tsukuiEmergence2025}, supplemented by an X-shaped structure -- reproduce the morphology across all JWST bands, with the geometry of each component shared across bands and only its flux normalization varying between them.

We assumed the disk to be perfectly edge-on, motivated by the nearly straight dust lane in the short-wavelength images. The axial ratio of the dust mask constructed below ($q_{\rm dust}=0.158$) implies a maximum deviation of $\lesssim 9^\circ$ from edge-on ($90^{\circ} -\cos^{-1}(q_{\rm dust})$); the true deviation is smaller because PSF blurring of the dust lane and the finite vertical thickness of the dust disk both make the dust lane appear rounder. Even this maximum deviation does not bias the disk scale radius \cite{tsukuiEmergence2025, yuThick2026} and thus does not affect the conclusions of this paper. Building on the edge-on disk–S\'ersic decomposition of ref. \cite{tsukuiEmergence2025}, we adopted a three-component model consisting of (1) an edge-on warped disk, (2) a S\'ersic component, and (3) an X-shaped structure.

\subsubsection*{Structural models}

\textbf{(1) An edge-on warped disk.} We modeled the stellar disk with a radially near-exponential profile and a vertical $\mathrm{sech}^2$ distribution. To allow deviations from a pure exponential profile, we introduced a shape parameter $k$ following \cite{dattathriDeprojection2024}. The stellar density distribution is:
\begin{equation}\label{eq:eq2}
    \nu(R,z)=\nu_0\exp(-(R/R_0)^k)\mathrm{sech}^2(z/(2z_0)).
\end{equation}
Here, $R_0$ is the scale radius, defined as the radius where the radial profile falls to $e^{-1}$ of its central value, and $z_0$ is the vertical scale height. The disk also exhibits a slight S-shaped distortion, which we modeled as a warp. The exact azimuthal orientation of the warp is uncertain and degenerate with the other warp parameters. We therefore assumed that the warp lies perpendicular to the line of sight. 

With this assumption, the warp is implemented by tilting each disk ring at radius $R$ about the line of sight axis by an angle $\theta$ \cite{binneyDisc2024}:
\begin{equation}\label{eq:eq3}
\theta(R)
= -h_{\textrm{max}}
\begin{cases}
0, 
    & R < R_{\textrm{warp}}-\Delta, \\[6pt]
\dfrac{R}{\,R_{\textrm{warp}}+\Delta\,},
    & R > R_{\textrm{warp}}+\Delta, \\[10pt]
\dfrac{1}{2}\!\left[1+\sin\!\left(\dfrac{\pi}{2\Delta}(R-R_{\textrm{warp}})\right)\right],
    & \text{otherwise}.
\end{cases}
\end{equation}
Here, $h_\textrm{max}$ is the warp maximum amplitude, $\Delta$ is the transition scale length, and $R_{\textrm{warp}}$ is the radius at which the warp occurs.

To construct the projected disk image, we generated $10^6$ particles with positions (R, $\phi$, z) drawn from the density distribution in Eq.~\ref{eq:eq2}. We then applied the disk warp transformation. The resulting particle distribution was projected onto the sky plane $(x, y)$ with the disk inclination (fixed to $90^\circ$) to obtain the surface-brightness distribution, which was subsequently convolved with the PSF. This particle-based approach is inspired by the \textsc{KinMS} code \cite{davisBlackhole2013}, which we modified for this analysis \cite{tsukuiTakafumi2912026}. Accurate recovery of the disk structural parameters was verified through comparison with the \textsc{IMFIT} code \cite{erwinIMFIT2015}.

The disk model has 8 free parameters: radial scale radius $R_0$, vertical scale height $z_0$, radial shape parameter $k$, total disk flux $F_{\rm disk}$, disk position angle $PA$, and the warp parameters ($h_\textrm{max}$, $\Delta$, and $R_{\mathrm{warp}}$). The warp is significant only at large disk radii and does not affect the modeling of the central S\'ersic component and the X-shaped structure. We verified that the measured disk parameters (scale height and radius) remain unchanged when the warp is turned off, with differences smaller than 4\%.

\textbf{(2) S\'ersic component}.
We modeled the centrally concentrated light with a two-dimensional S\'ersic profile \cite{sersicAtlas1968a} implemented in the \textsc{astropy} package \cite{astropycollaborationAstropy2022}. It has 4 free parameters: total flux $F_{S}$, half-light radius $R_{S}$, S\'ersic index $n_{S}$, and ellipticity $e_{S}$. The center and position angle of this component, as well as of the X-shaped structure described below, are fixed to those of the disk. We verified that allowing these parameters to vary freely instead does not change the measured structural properties.

\textbf{(3) X-shaped structure.} After subtracting the disk and S\'ersic components, the residual image reveals a clear X-shaped feature (Fig.~\ref{fig:Fig1}). We therefore included an additional component to reproduce this structure. 

We modeled the X-shaped component using a 2D S\'ersic profile modified with an $m=4$ Fourier mode \cite{pengDetailed2010}. The $m=4$ mode produces four symmetric ridges in the surface brightness distribution, reproducing the observed morphology. The surface brightness distribution is
\begin{equation}
I(x,y) =
I_e \exp\!\left\{
    -b_{n_{\mathrm{X}}}
    \left[
        \left(
            \frac{r'(x,y)}{R_X}
        \right)^{1/n_{\mathrm{X}}}
        - 1
    \right]
\right\}.
\end{equation}
Here, we adopt a sky coordinate $(x,y)$ centered on the component center $(x_0,y_0)$ where the x-axis is aligned with the disk major axis. The unperturbed elliptical radius is defined as
\begin{equation}
r(x,y)^2 = (x-x_0)^2 + \left(\frac{y-y_0}{1-e_{\mathrm{X}}}\right)^{\!2},
\end{equation}
where $e_{\mathrm{X}}$ is the ellipticity of the component.

The azimuthal angle from the major axis is
\begin{equation}
\phi = \arctan2\!\left(\frac{y-y_0}{1-e_{\mathrm{X}}},\,x-x_0\right),
\end{equation}
and the $m=4$ perturbation modifies the radius as
\begin{equation}
r'(x,y) = r(x,y)\left[1 + a_4 \cos(4\phi)\right],
\end{equation}
where $a_4$ controls the strength of the X-shaped distortion. $a_4=0$ reduces the equation to a standard 2D S\'ersic profile. The minor-to-major axis ratio (Fig.~\ref{fig:Fig1}C) is simply $b_X/a_X=1-e_{X}$. The opening angle of the X-shaped structure $\Phi_{\mathrm{ridge}}$, defined as the angle between the disk major axis and the first ridge, is related to the ellipticity by:
\begin{equation}\label{eq:eq8}
    \Phi_{\mathrm{ridge}}=\arctan(1-e_X)
\end{equation}
This formulation is similar to ref. \cite{savchenkoMeasuring2017}, but uses fewer parameters while still capturing the opening angle of the X-shaped structure via ellipticity, which is advantageous for data-limited high-redshift galaxies. The X-shaped structure has 5 free parameters: total flux $F_{X}$, effective radius $R_X$, S\'ersic index $n_{X}$, ellipticity $e_{X}$, and $m=4$ strength parameter $a_{4}$. The position angle of the major axis and central position are fixed to those of the disk. 

\begin{figure}
    \centering
    \includegraphics[width=1\linewidth]{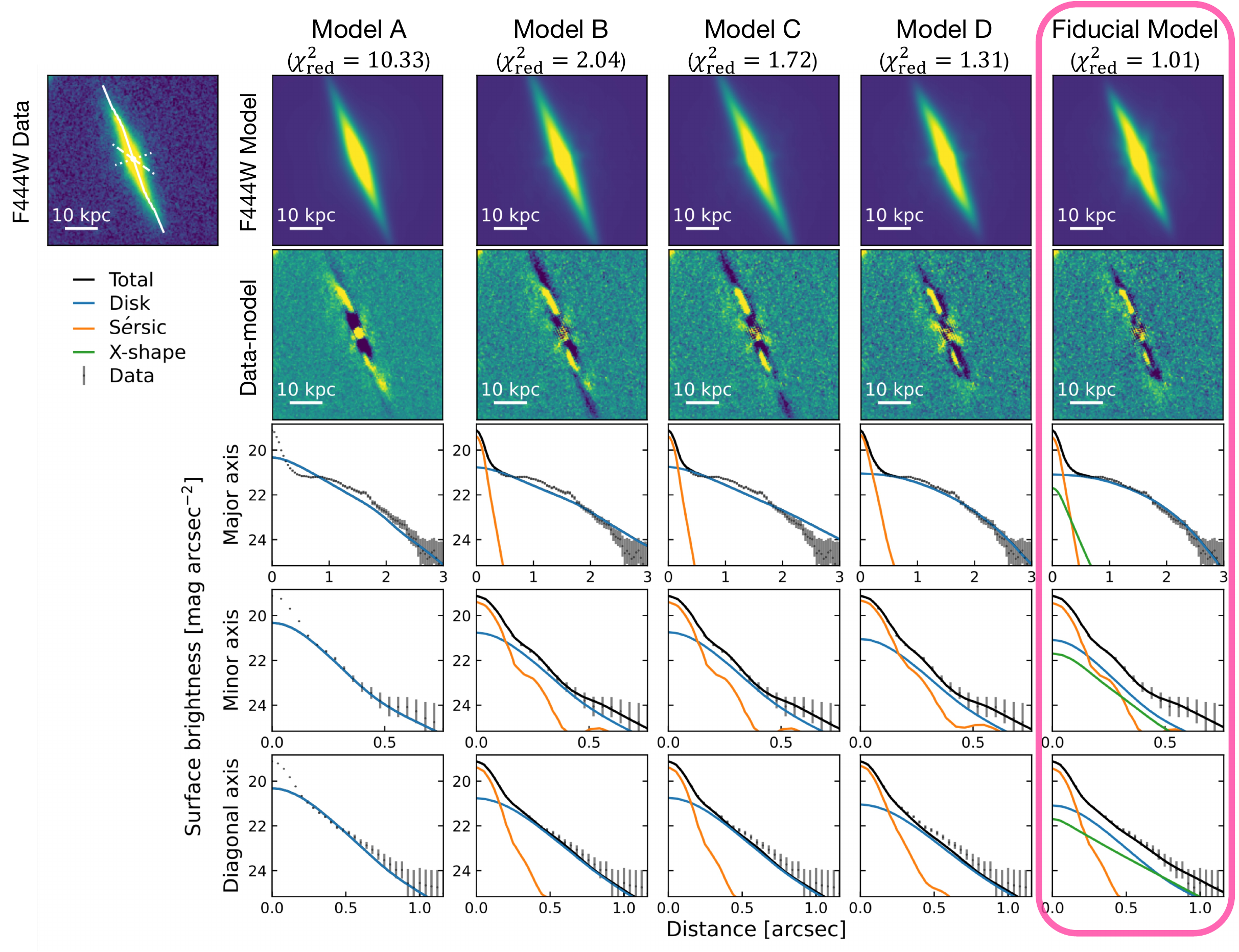}
    \caption{\textbf{Single-band structural model comparison for the F444W image.} The leftmost panel shows the observed image. Each column presents a different model (A-D and the fiducial model, from left to right). From top to bottom: best-fit model image, residual (data-model), and surface brightness profiles along the major axis (aligned with the best-fit disk position angle), minor axis, and diagonal directions (aligned with the best-fit X-shape ridges). The corresponding extraction paths are overplotted in white in the observed image. Black points with error bars denote the data, while lines show the total model (black) and its components: disk (blue), S\'ersic component (orange), and where included, the X-shaped component (green). Models A-D successively add complexity: a simple exponential disk (A), plus a S\'ersic component (B), plus disk warping (C), plus a deviation from a pure exponential disk (D). The fiducial model additionally includes an X-shaped component.}
    \label{fig:fig5}
\end{figure}

\subsubsection*{Fitting procedure}
To constrain the structural properties and SEDs of each structural component while avoiding dust attenuation, we followed the workflow below. \textrm{(1)} We started by modeling the least dust-affected F444W image (rest-frame 2.3 $\mu$m). \textrm{(2)} Using the best-fit structural parameters, we iteratively identified dust-attenuated regions in the shorter-wavelength images through sigma clipping while refitting the component normalizations, and constructed a dust mask. \textrm{(3)} We then applied this mask while fitting the model to all available JWST bands simultaneously. In this final step, the structural parameters are shared across bands, while the flux normalization of each component is independently optimized.

\subsubsection*{Likelihood and optimization}
The best-fitting parameters are obtained by minimizing the $\chi^2$ statistic
\begin{equation}
\chi^2(\theta) = \sum_b\sum_i \left(\frac{I_{b,i,\mathrm{data}} - I_{b,i,\mathrm{model}}(\theta)}{\sigma_{b,i}}\right)^2,
\end{equation}
where $b$ denotes the photometric band used in the fit and $i$ the unmasked pixel index. $I_{b,i,\mathrm{data}}$ and $I_{b,i,\mathrm{model}}$ are the observed and PSF-convolved model surface brightness values of pixel $i$ in band $b$, and $\sigma_{b,i}$ is the corresponding pixel uncertainty. We used the source pixels defined by the segmentation mask derived from the detection image, constructed by stacking the F277W, F356W, and F444W images \cite{valentinoAtlas2023}.%, while we later combine this mask with the dust mask when modelling the multicolour images.

The likelihood is written as: 
\begin{equation}
\mathcal{L}(\theta) = \exp\!\left(-\chi^2(\theta)/2\right).
\end{equation}

To ensure convergence in the high-dimensional parameter space, we first obtained initial parameter estimates through $\chi^2$ minimization using the Nelder-Mead simplex algorithm \cite{nelderSimplex1965} implemented in \textsc{scipy.optimize} \cite{gaoImplementing2012, virtanenSciPy2020}. These parameters are then used as starting points for Markov Chain Monte Carlo (MCMC) sampling with the \textsc{emcee} code to explore the posterior distribution and determine parameter uncertainties. We used uniform priors for all parameters in linear space. 

\subsubsection*{(1) Initial structural fit to the F444W image}
Fitting the structural model to the F444W image (rest-frame 2.3 $\mu$m) provides largely dust-unaffected estimates of the structural parameters for the three main components: the warped disk, the S\'ersic component, and the X-shaped structure. 

Figure~\ref{fig:fig5} compares, for the F444W image, the PSF-convolved best-fit models, data-model residuals and surface-brightness profiles along the major (aligned with the best-fit disk position angle), minor, and diagonal axes (aligned with the best-fit X-shape ridges) for a sequence of models that successively add complexity: a simple exponential disk (A); plus a S\'ersic component (B); plus disk warping (C); plus a disk shape parameter (D); and the fiducial model which additionally includes an X-shaped bulge to reproduce the X-shaped morphology seen in Model D residual. This comparison demonstrates that all components are required to reproduce the observed structure. Structural parameters of the three components in Model D are well constrained, with no significant degeneracies between components (see the corner plot of the multi-band fit in Fig.~\ref{fig:fig9}, shown later). The three components are therefore cleanly separable.

\begin{figure}
    \centering
    \includegraphics[width=0.375\linewidth]{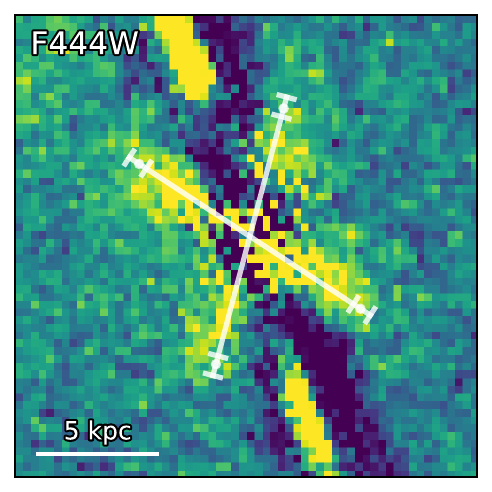}
    \includegraphics[width=0.375\linewidth]{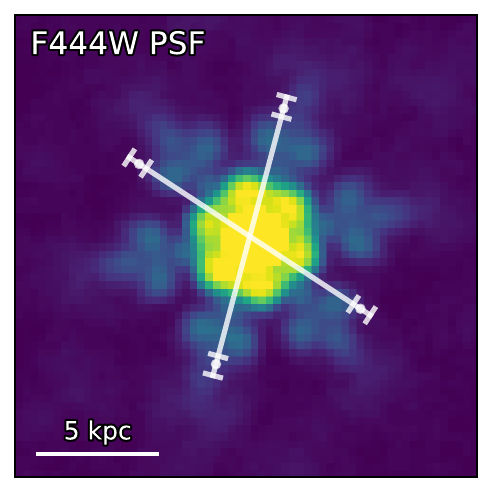}
    \caption{\textbf{Comparison between the observed X-shaped residual and the JWST/NIRCam PSF diffraction pattern of the F444W image.} Left: same as Fig.~\ref{fig:Fig1}C. The best-fit X-shaped structure (Fiducial model) is indicated by white lines. Right: the F444W PSF shown at the same physical scale and orientation.}
    \label{fig:fig6}
\end{figure}

The residual X-shaped morphology seen in Model D (Fig.~\ref{fig:fig5}D residual and Fig.~\ref{fig:Fig1}C) is intrinsic to the galaxy rather than a PSF artifact, as the diagonal surface-brightness profile of the fiducial model shows that the X-shaped component dominates at large radii and far exceeds the PSF-scattered light expected from the central compact component. In addition, the X-shaped morphology is inconsistent with the F444W PSF diffraction pattern (Fig.~\ref{fig:fig6}).

\subsubsection*{(2) Dust mask construction}
Because the dust attenuation produces localized flux deficits relative to the smooth stellar distribution, it can be identified from the residual in the short wavelength bands. We identified dust-affected regions by fitting the three structural components with the best-fit shapes obtained from the F444W image and allowing only their amplitudes to vary (3 parameters in total). The model was convolved with the point-spread function of the image to be fitted. We then updated the dust-lane mask based on residuals (model$-$data) exceeding $4\times\sigma_{\textrm{pix}}$, where $\sigma_{\textrm{pix}}$ is the pixel noise standard deviation. We repeated the fitting and mask-updating steps until the mask converged. Using different short-wavelength bands yields qualitatively consistent dust-lane masks. Throughout this paper, we adopted the mask constructed from F200W, which has a high signal-to-noise ratio over the disk region. We further expanded the mask conservatively by 1 pixel (0.04 arcsec).

\begin{figure}
    \centering
    \includegraphics[width=\linewidth]{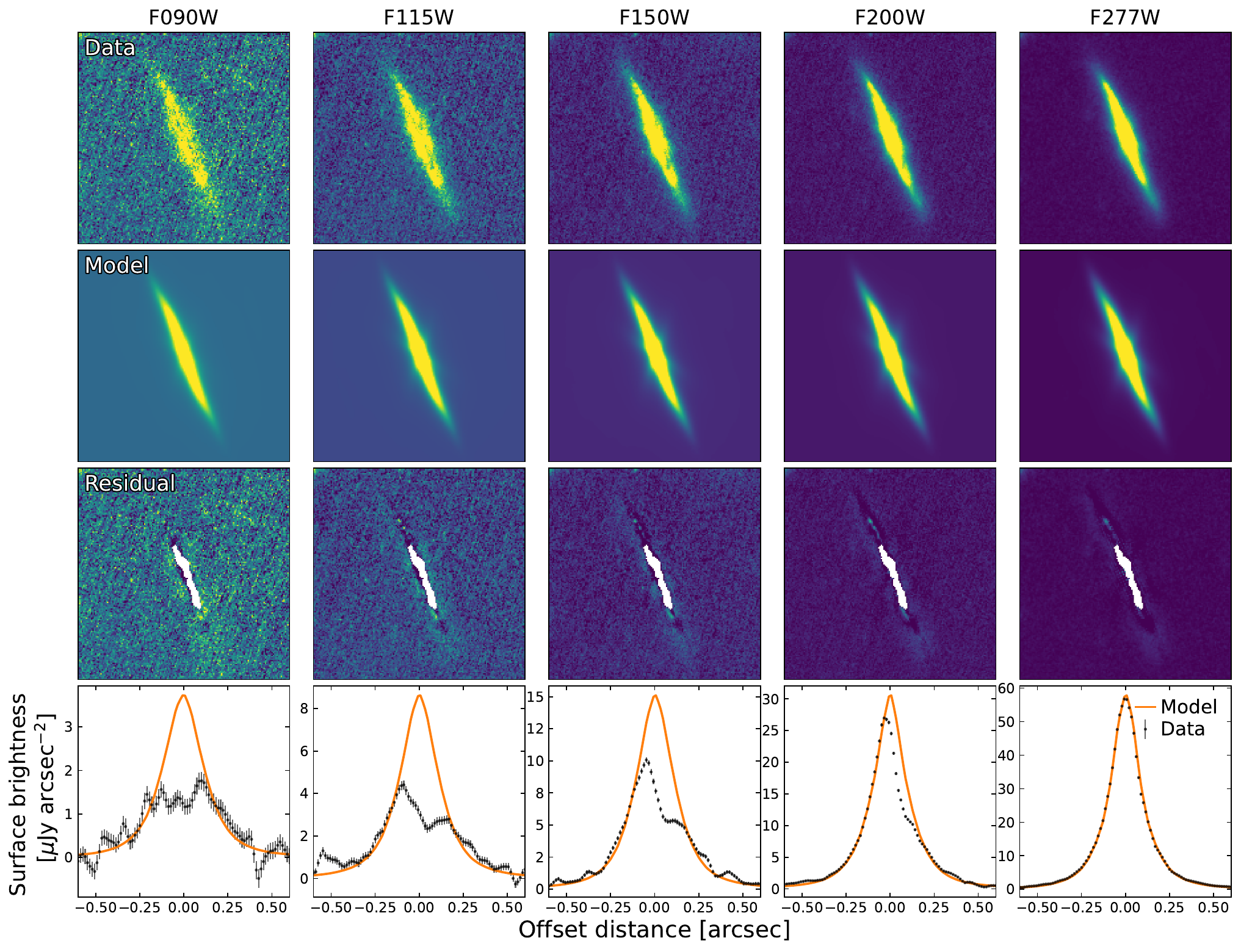}
    \caption{\textbf{Three-component model fits to short-wavelength bands (F090W, F115W, F150W, F200W, F277W, left to right) using the dust-lane mask.} We optimize only the flux normalization of each component, and fix all other structural parameters to the best-fit values from the F444W image (fiducial model in Fig.~\ref{fig:fig5}). From top to bottom, panels show the image, best-fit model, residual, and minor axis surface-brightness profile. White regions in the residual panels indicate the masked dust lane.}
    \label{fig:fig7}
\end{figure}

Figure~\ref{fig:fig7} compares the short wavelength data and the corresponding model, in which only the flux normalizations are optimized while the structural parameters are fixed to the best-fit values from the F444W image, with the dust lane mask applied. The residual images and minor-axis profiles show that the dust mask reduces biases in structural and flux measurements.

\begin{figure}
    \centering
    \includegraphics[width=\linewidth]{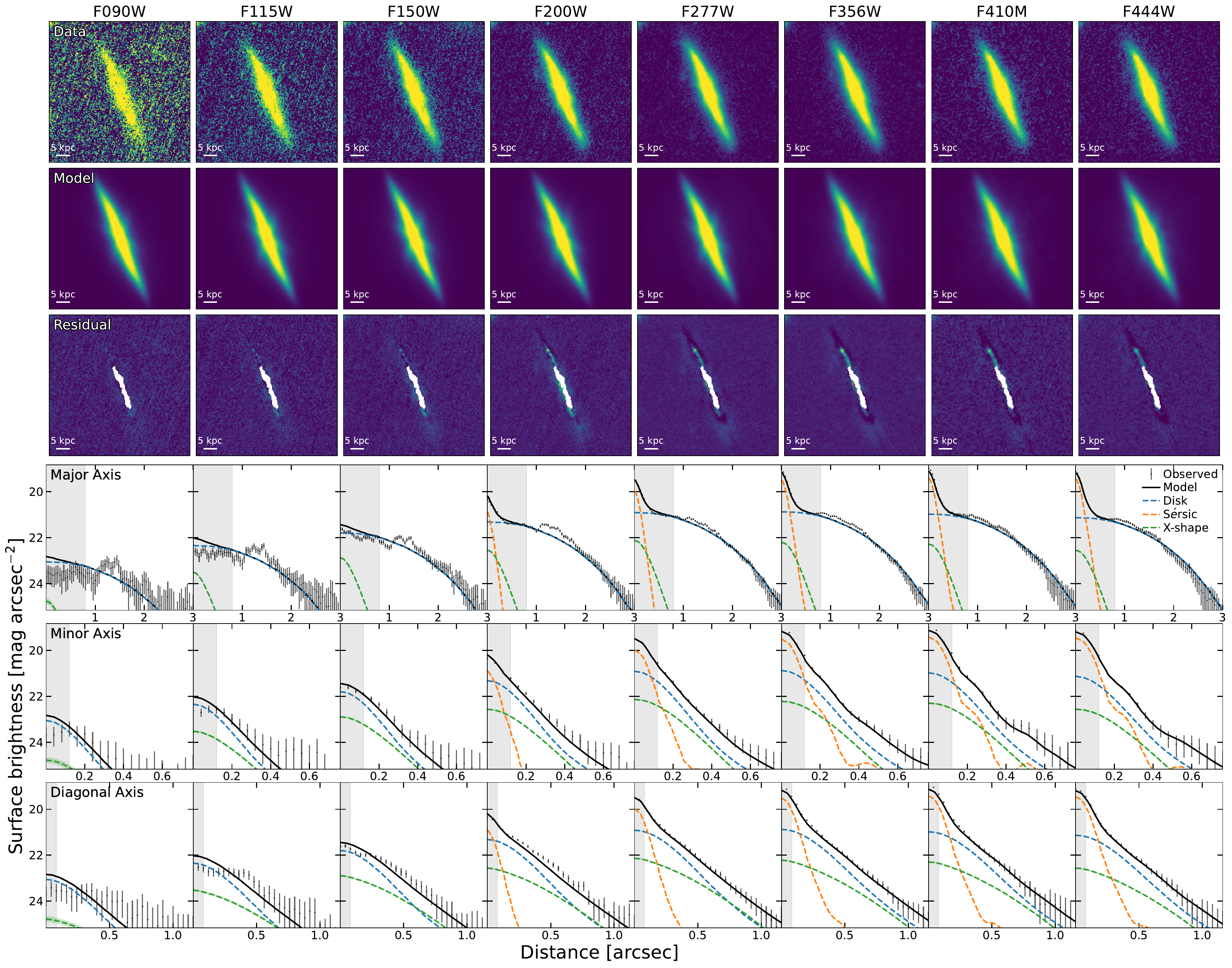}
    \caption{\textbf{The best-fit three-component model across JWST bands.} Columns show the F090W, F115W, F150W, F200W, F277W, F356W, F410M and F444W images (left to right). From top to bottom, panels show the data, best-fit model, residual (data-model), and surface brightness profiles along the major axis, minor axis, and diagonal directions (same as Fig.~\ref{fig:fig5}). Points with error bars denote the data, while lines show the total model (black) and its components: disk (blue), S\'ersic (orange), and X-shaped component (green). White regions in the residual panels indicate the masked dust lane, and gray shaded regions in the surface brightness profiles indicate the dust-masked radii.}
    \label{fig:fig8}
\end{figure}

\subsubsection*{(3) Final multi-band decomposition}
Using the dust mask, we fitted the model simultaneously to all available JWST bands (F090W, F115W, F150W, F200W, F277W, F356W, F410M, F444W). In total, we sampled the posterior distribution of 24 flux parameters (3 components across 8 bands) and 16 structural parameters shared across bands. Fig.~\ref{fig:fig8} demonstrates that the three-component model provides a good fit across all bands, with a varying contribution of the S\'ersic component, while the ratio between the X-shaped and disk components remains roughly constant. The best-fit model yields a reduced $\chi^2$ of 1.291 with 47,880 degrees of freedom, with the modest excess above unity potentially reflecting underestimated pixel errors due to correlated noise from image drizzling. To obtain conservative parameter uncertainties, we rescaled the error images upward by a constant factor so that the reduced $\chi^2$ equals unity prior to MCMC sampling.

\begin{figure}
    \centering
    \includegraphics[width=1\linewidth]{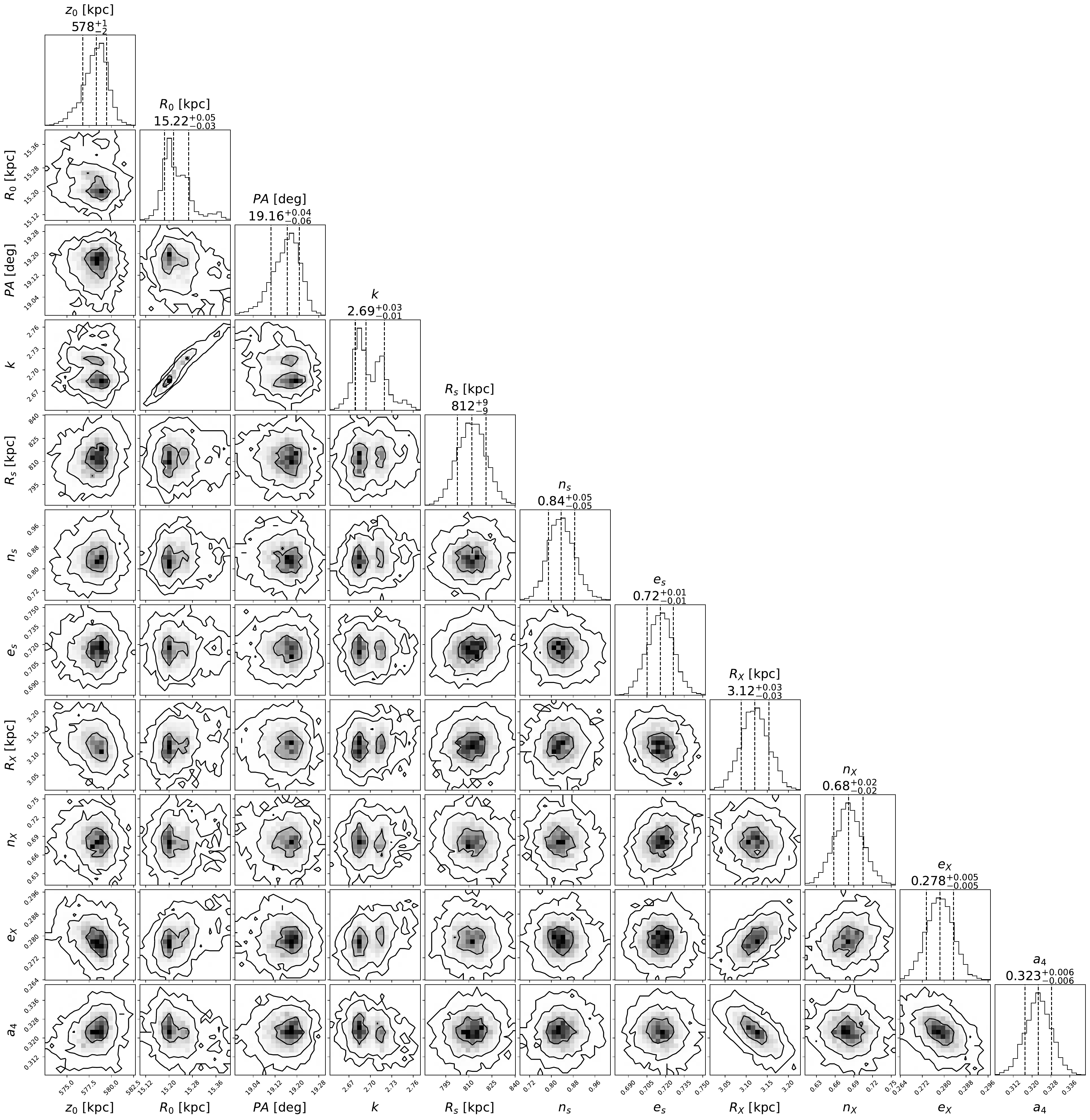}
    \caption{\textbf{Posterior probability distributions of the 11 structural parameters of our multi-color model.} For readability, we omit the fluxes of each component in each band (3 components $\times$ 8 bands; Table \ref{tab:tab4}) and the nuisance parameters: the central positions ($x_0$, $y_0$) and the warp parameters ($h_{\rm max}$, $\Delta$, and $R_{\rm warp}$), which are well constrained (Table \ref{tab:tab3}). Diagonal panels show the marginalized histograms of each parameter, with the median and 68\% credible interval shown at the top, while the other panels show the covariances between model parameters. The contours enclose 68\%, 95\%, and 99.7\% of the posterior probability.}
    \label{fig:fig9}
\end{figure}

The posterior distributions and the best-fit parameters are shown in Fig.~\ref{fig:fig9} and Table~\ref{tab:tab3}. The structural parameters obtained are consistent with those from the single-band F444W fit, and the multi-band fit additionally provides SED photometry for each component. The photometry of the X-shaped and disk components is constrained mainly by emission outside the dust-obscured midplane, whereas the compact central S\'ersic component is affected by high-opacity dust and unresolved dust geometry. Fluxes of the S\'ersic component in the three shortest-wavelength bands are consistent with zero, and we report 3$\sigma$ upper limits. We measured the photometry of each component by applying the same aperture used for the full galaxy SED to the PSF-convolved model images drawn from the posterior distribution. The resulting fluxes are summarized in Table~\ref{tab:tab4}. 

\begin{table}
\centering
\caption{\textbf{Best-fit structural parameters of the fiducial three-component model.} The central position $(x_0,y_0)$ and position angle $PA$ are shared by the three components. Reported values are the posterior median and 68\% confidence interval.}
\label{tab:tab3}
\begin{tabular}{l l c c}
\hline
\hline
Parameter & Unit & Symbol & Best-fit value \\
\hline

\multicolumn{4}{l}{\textit{Shared parameters}} \\
\hline
Central position in $x$ & pixel & $x_0$ & $0.45^{+0.01}_{-0.01}$ \\
Central position in $y$ & pixel & $y_0$ & $1.15^{+0.01}_{-0.01}$ \\
Position angle & deg & $PA$ & $19.16^{+0.04}_{-0.06}$ \\
\hline

\multicolumn{4}{l}{\textit{Disk component}} \\
\hline
Radial scale radius & kpc & $R_0$ & $15.22^{+0.05}_{-0.03}$ \\
Vertical scale height & pc & $z_0$ & $578^{+1}_{-2}$ \\
Radial shape parameter & -- & $k$ & $2.69^{+0.03}_{-0.01}$ \\
Warp amplitude & deg & $h_{\mathrm{max}}$ & $4.1^{+0.1}_{-0.1}$ \\
Warp transition scale & kpc & $\Delta$ & $6.3^{+0.3}_{-0.2}$ \\
Warp radius & kpc & $R_{\mathrm{warp}}$ & $13.9^{+0.1}_{-0.1}$ \\
Inclination (fixed) & deg & $i$ & $90$ \\
\hline

\multicolumn{4}{l}{\textit{S\'ersic component}} \\
\hline
Effective radius & pc & $R_S$ & $812^{+9}_{-9}$ \\
S\'ersic index & -- & $n_S$ & $0.84^{+0.05}_{-0.05}$ \\
Ellipticity & -- & $e_S$ & $0.72^{+0.01}_{-0.01}$ \\
\hline

\multicolumn{4}{l}{\textit{X-shaped component}} \\
\hline
Effective radius & kpc & $R_X$ & $3.12^{+0.03}_{-0.03}$ \\
S\'ersic index & -- & $n_X$ & $0.68^{+0.02}_{-0.02}$ \\
Ellipticity & -- & $e_X$ & $0.278^{+0.005}_{-0.005}$ \\
Fourier amplitude & -- & $a_4$ & $0.323^{+0.006}_{-0.006}$ \\
\hline

\end{tabular}
\end{table}

\begin{table}
\centering
\caption{\textbf{Best-fit fluxes of each component in each JWST band.} Values in units of $10\,\mathrm{nJy}$ are posterior median and 68\% credible interval. For the S\'ersic component in F090W, F115W, and F150W, we report 3$\sigma$ upper limits.}

\label{tab:tab4}
\begin{tabular}{lccc}
\hline
\hline
Band & $F_{\mathrm{disk}}$ $(10\,{\rm nJy})$ & $F_{S}$ $(10\,{\rm nJy})$ & $F_{X}$ $(10\,{\rm nJy})$ \\
\hline
F090W & $245.6^{+3.5}_{-4.1}$ & $<1.13$ (3$\sigma$) & $26.3^{+3.1}_{-3.0}$ \\
F115W & $467.1^{+5.0}_{-4.5}$ & $<0.65$ (3$\sigma$) & $81.9^{+3.0}_{-3.4}$ \\
F150W & $768.3^{+3.8}_{-4.6}$ & $<1.63$ (3$\sigma$) & $145.8^{+3.5}_{-3.1}$ \\
F200W & $1214.5^{+5.3}_{-5.2}$ & $55.1^{+2.6}_{-2.1}$ & $199.8^{+3.7}_{-3.5}$ \\
F277W & $1825.8^{+6.0}_{-5.9}$ & $180.8^{+4.0}_{-3.7}$ & $302.8^{+6.1}_{-4.9}$ \\
F356W & $1980.4^{+4.7}_{-5.7}$ & $333.3^{+4.2}_{-4.8}$ & $293.2^{+5.6}_{-5.5}$ \\
F410M & $1829.1^{+8.4}_{-7.2}$ & $397.3^{+4.5}_{-5.2}$ & $276.2^{+7.5}_{-6.5}$ \\
F444W & $1619.7^{+6.4}_{-6.9}$ & $408.5^{+5.0}_{-3.7}$ & $219.8^{+5.6}_{-5.3}$ \\
\hline
\end{tabular}
\end{table}

\subsubsection*{Best-fit structural model interpretation}
The major-axis profile of the fiducial model shows a S\'ersic component that is distinct from the extended disk (Fig.~\ref{fig:fig5}). The geometric properties of the S\'ersic component, including its size and axial ratio, are consistent with the nuclear stellar disks observed in nearby galaxies (see discussion in the main text). The extended disk has a nearly flat central radial profile, as captured by the disk shape parameter $k$. Similar flat profiles are also seen in nearby edge-on disk galaxies with X-shaped bulges, associated with the action of bar resonances in the disk \cite{BureauKband2006}. The flat disk profile, together with attenuation from the dust lane, may explain the slight negative residual near the center (Fig.~\ref{fig:fig5}) when the disk is modeled with a monotonically declining profile. 

Fig.~\ref{fig:Fig1}B-C and Fig.~\ref{fig:fig5} also reveal thin extended emission in the outer regions. The extension of this emission is consistent with the estimated bar size $11.8^{+5.5}_{-4.0}$ kpc, suggesting that it may trace the thin part of the bar \cite{weggStructure2015}, a thin stellar disk \cite{tsukuiEmergence2025}, or enhanced star formation at the bar end \cite{maedaStatistical2023}. We did not model this thin component, which appears both as a dust lane and as an extended emission in the outer disk (Fig.~\ref{fig:fig8}); doing so would require a substantially more complex model with a wavelength-dependent transition radius. The galaxy is dominated by the main disk component, and including the thin component does not significantly affect either the measured structural parameters \cite{tsukuiEmergence2025} or the main disk photometry measurements. 

The derived disk thickness is $z_0=578^{+1}_{-2}$ pc, and the scale radius is $R_0=15.22^{+0.05}_{-0.03}$ kpc (Tab.~\ref{tab:tab3}). The $R_0$ value can be converted to an effective radius enclosing half of the total flux, $R_{\rm d, eff}=11.28^{+0.01}_{-0.01}$ kpc. For comparison with the literature, we also derive an exponential disk scale radius of $R_{\rm d, exp}=6.13^{+0.02}_{-0.01}$ kpc from Model C which uses an exponential profile with $k=1$ (Fig.~\ref{fig:fig5}).

\subsubsection*{SED modeling}
We modeled the spectral energy distribution of the galaxy (Table~\ref{tab:tab2}) and its structurally decomposed components (the disk and X-shaped components, Table \ref{tab:tab4}) using \textsc{Prospector} \cite{johnsonStellar2021} (version 1.4.1). The central S\'ersic component, associated with the nuclear stellar disk, is strongly affected by dust attenuation, and its inferred stellar population depends sensitively on the assumed extinction curve and dust geometry \cite{nishiyamaInterstellar2009, gonzalezReddening2012}. We therefore restricted our SED analysis to the X-shaped and disk components. The model represents the observed SED as a superposition of stellar population spectra parametrized by the SFH and stellar metallicity, modified by dust attenuation. For the decomposed components, we limited the fitting to the NIRCam bands, where the structural decomposition is feasible. 

\begin{figure}
    \centering
    \includegraphics[width=0.75\linewidth]{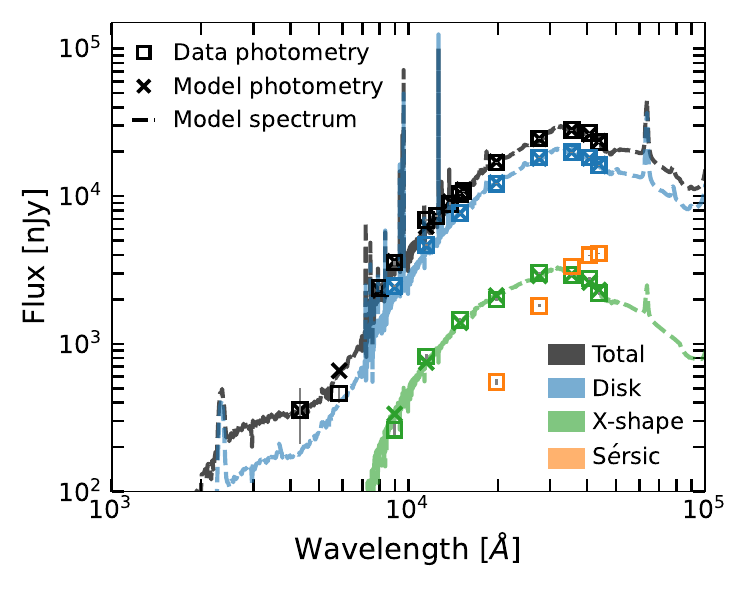}
    \caption{\textbf{Observed and model spectral energy distribution of UDS 12999.} Squares show the aperture photometry, and crosses show the best-fit \textsc{Prospector} model photometry. The dashed curve shows the best-fit model spectrum. The symbol colors distinguish the total galaxy (black) and the individual components: disk (blue), X-shaped structure (green), and S\'ersic component (orange).}
    \label{fig:fig10}
\end{figure}

We used the spectroscopic redshift derived from the NIRSpec spectrum ($z = 0.92331$). We adopted the standard non-parametric SFH \cite{lejaOlder2019} with the default seven age bins initialized using \textsc{adjust\_continuity\_agebins} in \textsc{Prospector} code. The youngest age bins cover 0-30 and 30-100 Myr; the oldest age bin spans 25\% of the age of the Universe (6.2 Gyr at the redshift of 0.92331); and the remaining bins are spaced evenly in logarithmic time. We assumed a Chabrier initial mass function \cite{chabrierGalactic2003}, the MIST isochrones \cite{DotterMesa2016, ChoiMesa2016}, the MILES spectral library \cite{sanchez-blazquezMILES2006}, and the flexible dust attenuation law by ref. \cite{kriekDust2013}. 
We adopted uniform priors for $\log M_*$ and $\log Z_*$ in the ranges [8, 12] and [-2.5, 1.5], respectively. For dust attenuation, we used the identical prior choice as ref. \cite{wangUNCOVER2024}. Figure~\ref{fig:fig10} shows the SED data points for the galaxy, disk, X-shaped component and S\'ersic component, together with the best-fit SED models.  

\subsubsection*{Structural measurement of the X-shaped structure}
\begin{figure}
    \centering
    \includegraphics[width=0.6\linewidth]{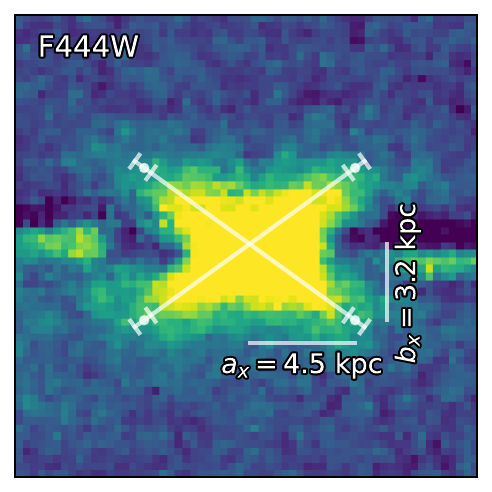}
    \caption{\textbf{The X-shaped component isolated from the F444W image.} The F444W image after subtraction of the disk and S\'ersic components of the fiducial best-fit model (disk, S\'ersic and X-shaped structures; Fig.~\ref{fig:fig5}), rotated so that the disk major axis is aligned with the x-axis. The best-fit opening angle and measured length of the X-shaped structure are overlaid.}
    \label{fig:fig11}
\end{figure}

We measured the length of the X-shaped structure following ref. \cite{savchenkoMeasuring2017}. Using the fiducial model (Fig.~\ref{fig:fig5}), we subtracted the S\'ersic and disk components from the F444W image to isolate the X-shaped component, and extracted surface brightness profiles along the four ridges defined by the best-fit opening angle. For each ridge, we defined the length as the maximum radius at which the signal remains above a given noise threshold. Because the profile declines more steeply than the noise, the result is only weakly sensitive to this choice: we adopted the value at $2.5\sigma$ as a central estimate and the values at $3\sigma$ and $2\sigma$ as the statistical uncertainty range. All quantities are averaged over the four ridges. This yields a length of $l=5.4^{+0.5}_{-0.3}$ kpc, with major and minor axis lengths $a_X = l \cos(\Phi_{\rm ridge}) = 4.5^{+0.4}_{-0.3}$ kpc and $b_X = l \sin(\Phi_{\rm ridge}) = 3.2^{+0.3}_{-0.2}$ kpc, in agreement with values obtained by the visual-inspection method of ref. \cite{laurikainenBarlenses2017} (Fig.~\ref{fig:Fig1}C). The measured length does not strongly depend on the band used. Figure~\ref{fig:fig11} compares the X-shaped residual and the measured X-shape size. 

%%%%%%%%%%% CAPTIONS FOR OTHER SUPPLEMENTARY FILES %%%%%%%%%%

\clearpage % Clear all remaining figures and tables then start a new page

\end{document}